\documentclass{ptephy}

\preprintnumber{XXXX-XXXX}

\usepackage{cite}

\begin{document}

\title{Performance of prototype Dual Gain Multilayer Thick GEM with high-intensity heavy-ion beam injections in low-pressure hydrogen gas}

\author{\name{\fname{Chihiro} \surname{Iwamoto}}{1\dagger*}, 
\name{\fname{Shinsuke} \surname{Ota}}{1,2}, 
\name{\fname{Reiko} \surname{Kojima}}{1}, 
\name{\fname{Hiroshi} \surname{Tokieda}}{1}, 
\name{\fname{Seiya} \surname{Hayakawa}}{1}, 
\name{\fname{Yutaka} \surname{Mizoi}}{3},
\name{\fname{Taku} \surname{Gunji}}{1},
\name{\fname{Hidetoshi} \surname{Yamaguchi}}{1},
\name{\fname{Nobuaki} \surname{Imai}}{1},
\name{\fname{Masanori} \surname{Dozono}}{1},
\name{\fname{Ryo} \surname{Nakajima}}{1},
\name{\fname{Olga} \surname{Beliuskina}}{1},
\name{\fname{Shin'ichiro} \surname{Michimasa}}{1},
\name{\fname{Rin} \surname{Yokoyama}}{1},
\name{\fname{Keita} \surname{Kawata}}{1},
\name{\fname{Daisuke} \surname{Suzuki}}{4},
\name{\fname{Tadaaki} \surname{Isobe}}{4},
\name{\fname{Juzo} \surname{Zenihiro}}{4},
\name{\fname{Yohei} \surname{Matsuda}}{5,6},
\name{\fname{Jun} \surname{Okamoto}}{5},
\name{\fname{Tetsuya} \surname{Murakami}}{7},
\name{\fname{Eiichi} \surname{Takada}}{8}
\thanks{Present address: Neutron Beam Technology Team, RIKEN Center for Advanced Photonics, RIKEN.}}

\address{\affil{1}{Center for Nuclear Study, University of Tokyo, Wako Saitama Japan}
\affil{2}{Research Center for Nuclear Physics, Osaka University, Ibaraki Osaka Japan}
\affil{3}{Osaka Electro-Communication University, Neyagawa Osaka Japan}
\affil{4}{RIKEN Nishina Center, RIKEN, Wako Saitama Japan}
\affil{5}{CYRIC, Tohoku University, Aoba-ku Sendai Miyagi Japan}
\affil{6}{Konan University, Higashinada-ku Kobe Hyogo Japan}
\affil{7}{Kyoto University, Sakyo-ku Kyoto Japan}
\affil{8}{National Institutes for Quantum and Radiological Science and Technology, Inage Chiba Japan}
\email{iwamoto@cns.s.u-tokyo.ac.jp}}

\begin{abstract}
A prototype Dual Gain Multilayer Thick Gas Electron Multilyer (DG-M-THGEM) 
with an active area of 10~cm $\times$ 10~cm was manufactured 
aiming at the production of a large-volume active-target time projection chamber which can work under the condition of high-intensity heavy-ion beam injections.
The DG-M-THGEM has a alternating structure of electrodes and insulators. 
Effective gas gains of two regions, which are called beam and recoil regions, are separately controlled.
Performance of the prototype DG-M-THGEM in hydrogen gas at a pressure of 40~kPa was evaluated.
Irradiating a $^{132}$Xe beam, an effective gas gain lower than 100 with a charge resolution of 3\% was achieved in the beam region 
while the effective gas gain of 2000 was maintained in the recoil region.
Position distributions of measured charges along the beam axis were investigated in order to evaluate gain uniformity in the high intensity beam injection. 
The gain shift was estimated by simulations considering space charges in the drift region. 
The gain shift was suppressed within 3\% even at the beam intensity of $2.5 \times 10^{6}$ particles per second. 
\end{abstract}

\subjectindex{H11, C30}

\maketitle


\section{Introduction}\label{sec1}
Gaseous active targets based on time projection chambers (TPCs)~\cite{Maya, ATTPC, ACTAR-TPC, MSTPC-1, MSTPC-2, MAIKo} 
as three dimensional tracking detector have been widely developed to perform experimental studies in inverse kinematics 
in various accelerator facilities.  
Gaseous active target plays an important role to measure forward-angle inelastic scattering of medium-heavy, especially unstable, nuclei in inverse kinematics.
The reaction vertex can be determined by reconstructing trajectories both beam and recoil particles simultaneously measured by the active target.
Inelastic scattering measurements involved by light nuclei such as proton, 
deuteron and $\alpha$ particles at incident energies of 100 - 300~MeV/nucleon are well known methods 
to determine transition strength of isovector-dipole and isoscalar-monopole states~\cite{Pb-pinl_PDR, Sn-pinl_PDR2, Sn-pinl_PDR4, Zr-pinl_PDR,  PbSn-dinl_GMR, Pb-dinl_GMR, Pb-ainl_ISGR, Sm-ainl_ISGR, Mo-ainl_ISGR, Sn-ainl_GMR}.
As radioactive isotope (RI) beams of medium-heavy nuclei, such as $^{132}$Sn region, 
with higher-intensity over $1 \times 10^{5}$ particles per second (pps) are becoming available 
at the energies for inverse-kinematics measurements of inelastic scatterings, 
the needs of active targets that can work under the condition of such high-intensity beam irradiation are increasing.

Strong ionization induced by heavy-ion and high-intensity beams are discussed in Ref.~\cite{MayaMask-2}, 
where the risk of sparks and malfunctioning of the detector are pointed out. 
The large number of the ionized electrons also inclease backflow ions from multiplication part. 
The backflow ions distort electric field in drift region and reduce the accuracy of the trajectory deduction.
In order to reduce these effects, the electrons and ions emerging inside an active area has to be controlled, especially along the beam trajectories. 
For example, the electrons and ions created by the beam particles are isolated from the drift region by placing plates in
MAYA~\cite{MayaMask-1, MayaMask-2}, and by surrounding the beam axis with wire rings in TACTIC~\cite{TACTICShield}. 
They can accept the high-intensity beams up to $5 \times 10^{7}$~pps, but they are not sensitive to the beam tracks. 
We tried to reduce electrons by means of a mesh grid covering the gas multiplication part~\cite{CATMeshGrid} in our active target, called CAT-S~\cite{SOtaCATS}.
The CAT-S with the mesh grid could detect beam trajectories, but the charge resolution was not better than 10\%, 
which was required to achieve the aiming position resolution of 1~mm.

For improving charge resolution of the CAT-S, we developed a Dual-Gain Thick Gas Electron Multiplier (DG-THGEM)~\cite{SOtaCATS}. 
The electrodes of the DG-THGEM at both sides are segmented to have individual gas gains for beam and recoil regions.
Independent gas gains of $1 \times 10^{2}$ and $5 \times 10^{3}$ were realized for beam and recoil regions, respectively.
The charge resolution better than 10\% for the beam region was achieved.
CAT-S with DG-THGEM could work under the condition of irradiation of RI beams including $^{132}$Sn at the high intensity of $3.5 \times 10^{5}$~pps~\cite{RIBF113}.

Now, we are developing a new active target TPC, named CAT-M, which has an active area of 30~cm~$\times$~30~cm larger  
than CAT-S with an active area of 10~cm~$\times$~10~cm, 
to increase the target thickness and the acceptance for reaction events. 
It is necessary to employ sufficient thickness of GEMs for enlarging active area in CAT-M, 
because its self-weight and Coulomb force acting between electrodes may deform itself.
A Multilayer THGEM (M-THGEM) developed by Cortesi et al.~\cite{NSCLMTHGEM} has an alternating structure of the electrodes and insulators.
This has sufficient thickness to resist the deformation.

We designed a new multilayer THGEM with a capability of the dual gain. We call it Dual Gain Multilayer Thick GEM (DG-M-THGEM).
In the present work, the prototype DG-M-THGEM with a same active area with CAT-S of 10~cm~$\times$~10~cm was used 
to investigate the gain stability and charge responses with high-intensity heavy-ion beam injection. 
The structure of the prototype DG-M-THGEM is described in Section~2. 
The measured effective gas gain and charge resolution of the DG-M-THGEM using a heavy-ion beam and the gain shift as a function of the beam intensity are shown in Section~3.
The Summary is given in Section~4.

\section{Structure of prototype DG-M-THGEM}\label{sec2}
\begin{figure}[h]
	\begin{center}
		\includegraphics[width=1.0\linewidth, bb=0 0 2384 2384]{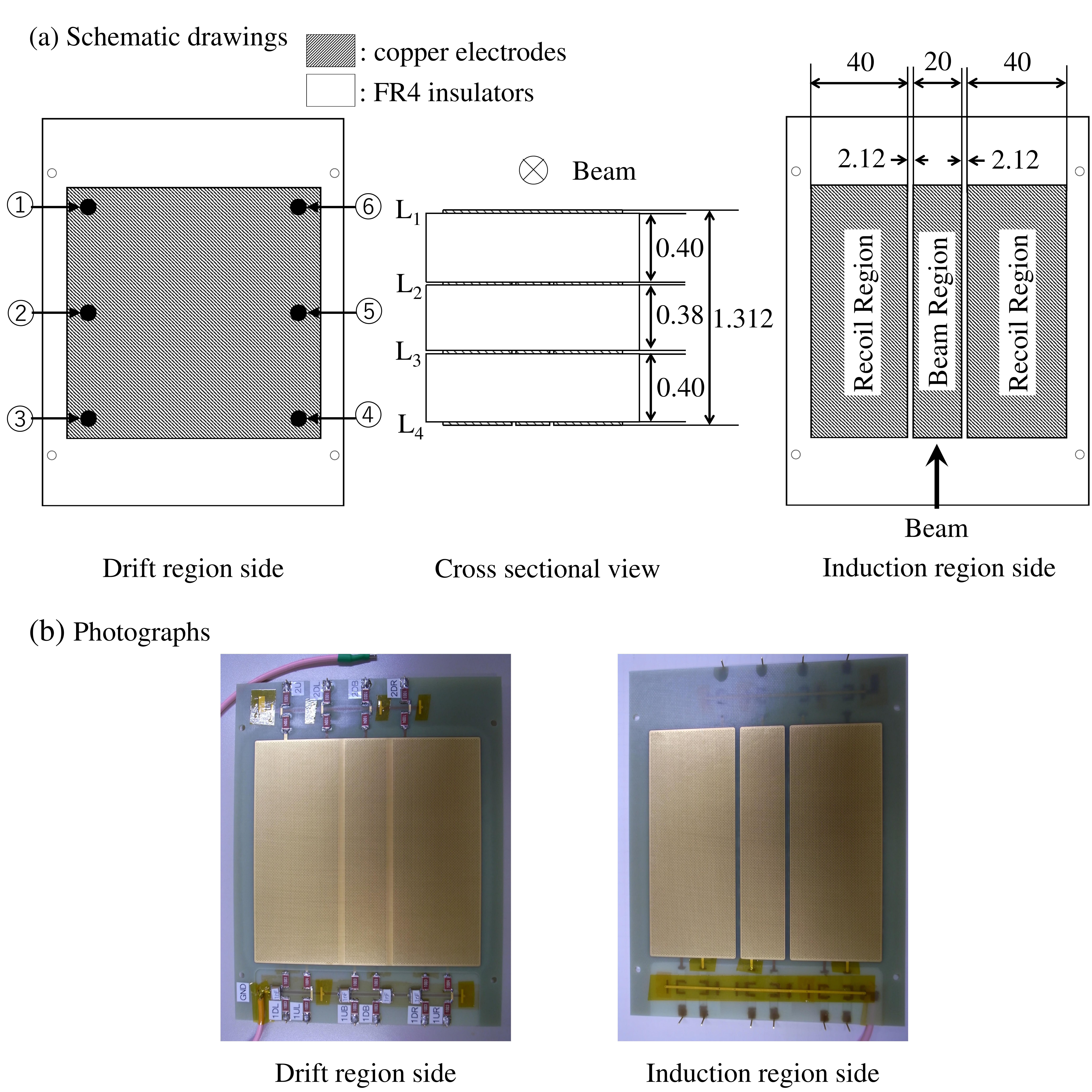}
		\caption{(color online) (a):~Schematic views of prototype DG-M-THGEM. Hatched regions are copper electrodes. Plain regions are FR4 insulators. Dimensional values are nominal in mm. Solid circles in the view from drift region side are indicated positions where thickness of DG-M-THGEM was measured. The number corresponds to the measured point in Table~\ref{actthickness}. Four electrodes are denoted by L$_{1}$, L$_{2}$ ,
L$_{3}$ and L$_{4}$ in the cross sectional view. The electrodes L$_{2}$ , L$_{3}$ and L$_{4}$ are divided into one beam region at the center and two recoil regions at both sides as shown in the view from induction region side. (b):~Photographs of the prototype DG-M-THGEM.}\label{DGMTHGEM}
	\end{center}
\end{figure}

\begin{figure}[h]
	\begin{center}
		\includegraphics[width=1.0\linewidth,bb=0 0 2007 1417]{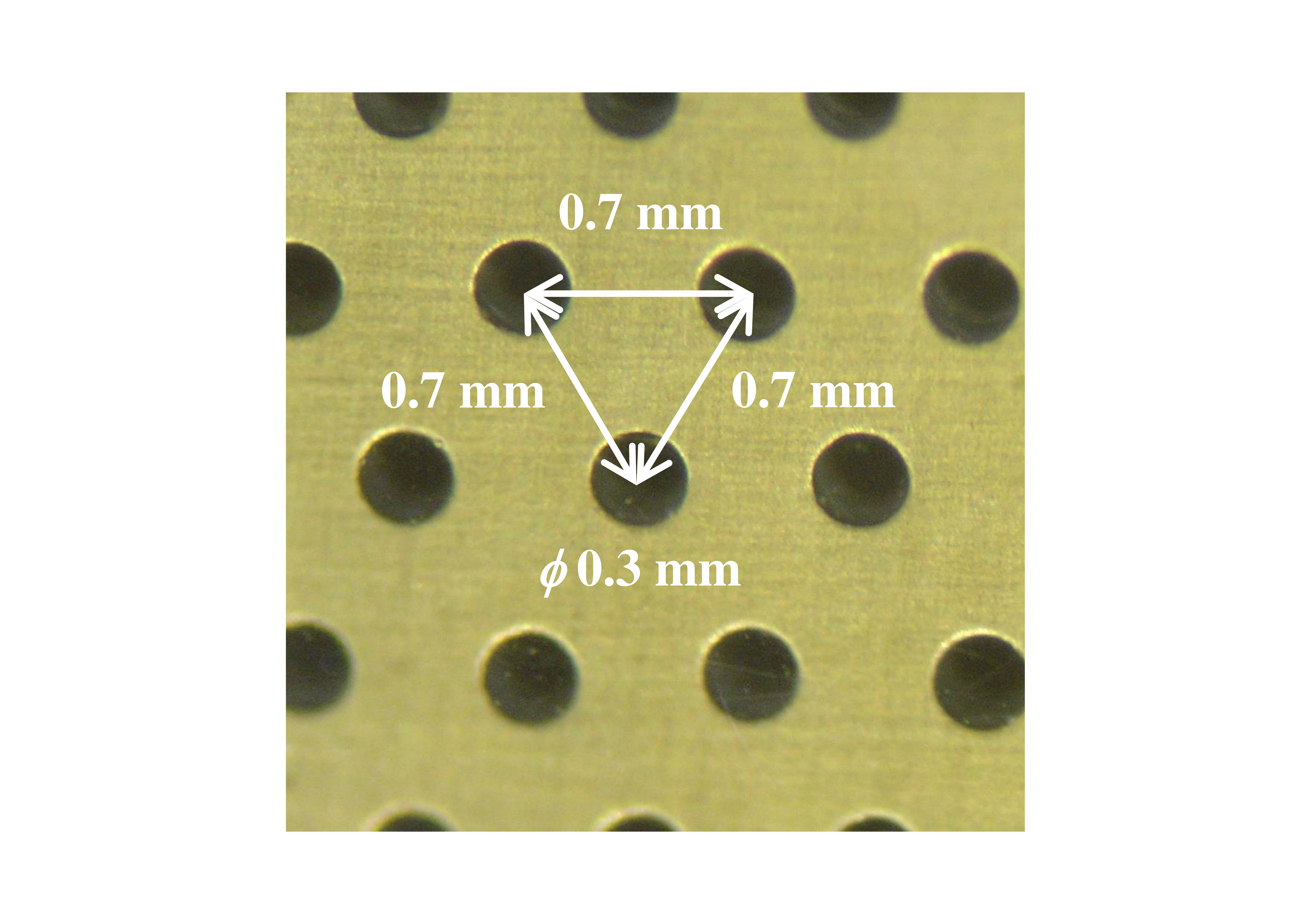}
		\caption{(color online) Photograph of DG-M-THGEM hole array pattern. Diameter and pitch of the holes are 0.3~mm and 0.7~mm, respectively.}\label{HoleSize}
	\end{center}
\end{figure}

Figure~\ref{DGMTHGEM} shows schematic drawings and photographs of the prototype DG-M-THGEM. 
It was designed by ourselves and produced by REPIC Co. Ltd, Japan.
The prototype DG-M-THGEM has an alternating structure of four sheets of electrodes and three plate of insulators.
Hereafter, four electrodes are denoted by L$_{1}$, L$_{2}$, L$_{3}$ and L$_{4}$.
The present production process is as following; 
first, two substrates with electrode copper layers on both sides were produced, i.e., one has the electrodes L$_{1}$ and L$_{2}$, and the other has the electordes L$_{3}$ and L$_{4}$. 
The thicknesses of copper electrode and FR4 insulator are 0.032~mm and 0.40~mm respectively. 
Then a 0.3-mm thick FR4 plate was sandwiched by these two substrates and bonded with 0.04-mm-thick epoxy-resin glue. 
The actual thickness of the middle substrate is 0.38~mm including the glue thickness.
And finally, GEM holes were drilled piercing from L$_{1}$ to L$_{4}$.
The geometry pattern of GEM holes is shown in Fig.~\ref{HoleSize}.
The diameter and the pitch of the hole are 0.3~mm and  0.7~mm, respectively.
A nominal total thickness of the prototype DG-M-THGEM is 1.312~mm. 
The actual thickness was measured at six points which are indicated in the view of the drift region side of Fig.~\ref{DGMTHGEM}~(a).
Table~\ref{actthickness} shows the measured thickness of each point with a caliper.
The value of each point is an average of the three measurements.
The average thickness at the six measured points is 1.298~mm and the maximum difference among the measured points is 0.002~mm.
The size of the substrate is 124~mm~$\times$~160~mm, and the area of the active region is approximately 100~mm~$\times$~100~mm. 
The electrodes L$_{1}$ and L$_{4}$ are closest to the drift and the induction regions, respectively.
The electrodes L$_{2}$, L$_{3}$ and L$_{4}$ are divided into three parts. 
The center one is called ``beam region'', and the two side ones are called ``recoil region''.
The beam region is 20-mm wide to cover the envelope of the beam, and each recoil region is 40-mm wide. 
There are 2.12~mm width gaps between the beam and recoil regions. 
The electrode L$_{1}$ is not segmented in order to avoid the charge up of the insulator.
As shown in photograph of Fig.~\ref{DGMTHGEM}~(b), protective resistors and stabilization capacitors are soldered on the insulator board.
\begin{table}[h]
\caption{Measured thicknesses of prototype DG-M-THGEM at positions indicated in Fig.~\ref{DGMTHGEM}. Each value is an average of the three measurements.}
\label{actthickness}
\centering
\begin{tabular}{c c}
\hline
measured position & actual thickness [mm] \\ 
\hline
1 & 1.301 $\pm$ 0.001 \\
2 & 1.290 $\pm$ 0.002 \\
3 & 1.294 $\pm$ 0.002 \\
4 & 1.303 $\pm$ 0.001 \\
5 & 1.302 $\pm$ 0.001 \\
6 & 1.297 $\pm$ 0.002 \\
\hline
\end{tabular}
\end{table}
\section{Performances of prototype DG-M-THGEM} \label{sec3}
For performance evaluations of the prototype DG-M-THGEM, two experiments were carried out.
First, effective gas gain of the prototype DG-M-THGEM was measured with various bias settings 
using an $\alpha$ source of $^{241}$Am in a test bench chamber. 
Second, the prototype DG-M-THGEM was installed in the CAT-S chamber. 
The effective-gas-gain stabilities and charge resolutions in the beam region were evaluated using a heavy-ion beam.
In both case, hydrogen gas with a purity of 99.99\% at the pressure of 40~kPa was filled.

\subsection{Effective gas gain evaluated using $^{241}$Am $\alpha$ source}\label{subsec3}
\begin{figure}[h]
	\begin{center}
		\includegraphics[width=1.0\linewidth, bb=0 0 2384 1417]{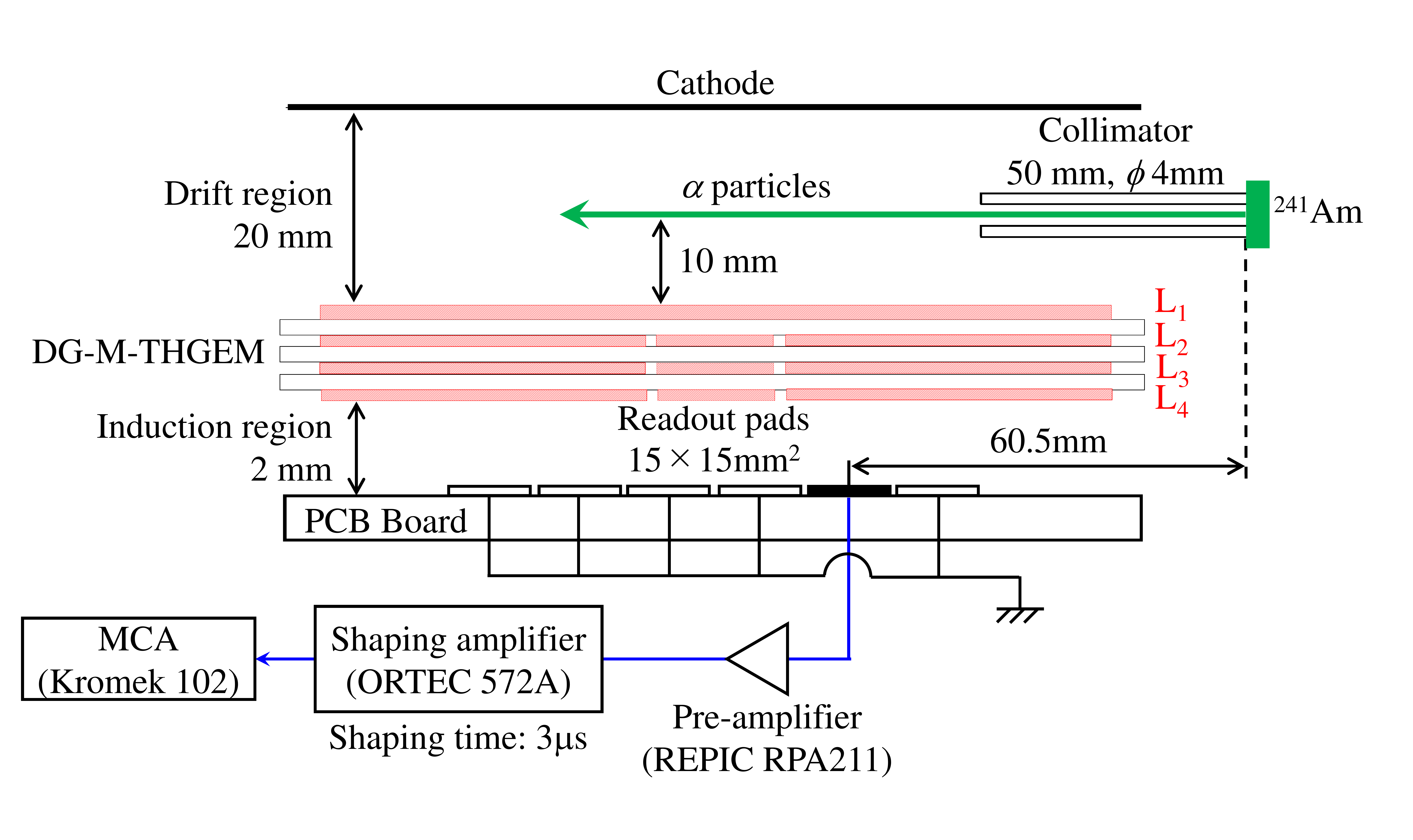}
		\caption{(color online) Schematic view of configuration in a test chamber and data acquisition system for effective gas gain measurement with $\alpha$ source $^{241}$Am. 
		Position relation of DG-M-THGEM to cathode, the $\alpha$ source, readout board  are shown.  
		Note that the dimensions of each component and each distance are not scaled.
		Rectangles on PCB board are segmented electrodes. Filled rectangle indicates readout pad while unfilled rectangles indicate the grounded electrodes. 
		The readout pad was located below recoil region of DG-M-THGEM and distance from the $\alpha$ source was 60.5~mm.
		Signals from the pad were measured by a multi-channel analyzer after amplified by a pre-amplifier and shaping amplifier. }\label{ExpSetOffline}
	\end{center}
\end{figure}

A schematic view of the experimental setup to measure the effective gas gain using the $^{241}$Am $\alpha$ source (0.98~kBq at the time of this experiment) is shown in Fig.~\ref{ExpSetOffline}. 
The test bench was installed in a cylindrical chamber with the inner diameter of 24~cm and the height of 12~cm.
In the chamber, the hydrogen gas flowed at the rate of approximately 100~cm$^{3}$/min. 
The gas pressure was monitored at the inlet and the outlet position of the chamber with differential pressure gauges  with the accuracy of 0.1~kPa. 
Oxygen concentration in the chamber was also monitored periodically at the outlet position and 0.01\% lower limit of the monitor or less of the oxygen concentration was kept. 
A cathode plate was placed above L$_{1}$ with distance of 20~mm from surface of L$_{1}$. 
Drift field for electrons was formed by L$_{1}$ and the cathode plate.
The $\alpha$ particles, which were collimated by a PTFE collimator with the length of 50~mm and the inner diameter of 4~mm, were injected to the drift field.
The $\alpha$ particles were injected to the recoil region and passed through 10~mm above L$_{1}$.  
A readout board was placed 2~mm below the surface of L$_{4}$.
On the readout board, 36 square-shaped copper pads, each of which has a 15~mm~$\times$~15~mm area, 
were arranged in 6~rows and 6~columns with interval gaps of 1~mm. 
One pad, hereafter called readout pad, shown as filled rectangle in Fig.~\ref{ExpSetOffline} (a) was connected to a readout circuit. 
The other pads were all grounded. 
The distance between the center of the readout pad and the surface of the $\alpha$ source was 60.5~mm.
The energy deposit  of $\alpha$ particle passing through the area corresponding to the readout pad was calculated to be 115~keV by LISE++~\cite{LISE++}. 
The collected electrons on the readout pad were integrated with a charge-sensitive preamplifier, REPIC RPA-211. 
Conversion gain was modified to be 400~mV/pC and a time-constant of 80~ns. 
The output signal from the preamplifier was pulse-shaped by a shaping amplifier, ORTEC 572A, 
and its pulse height was recorded by a multi-channel analyzer, Kromek 102 product of Kromek Group plc. 
In order to convert the pulse height to the absolute charge value, the circuit system was calibrated by a pulser module and charge injector (capacitance).

\begin{figure}[h]
	\begin{center}
		\includegraphics[width=1.0\linewidth, bb=0 0 2205 1559]{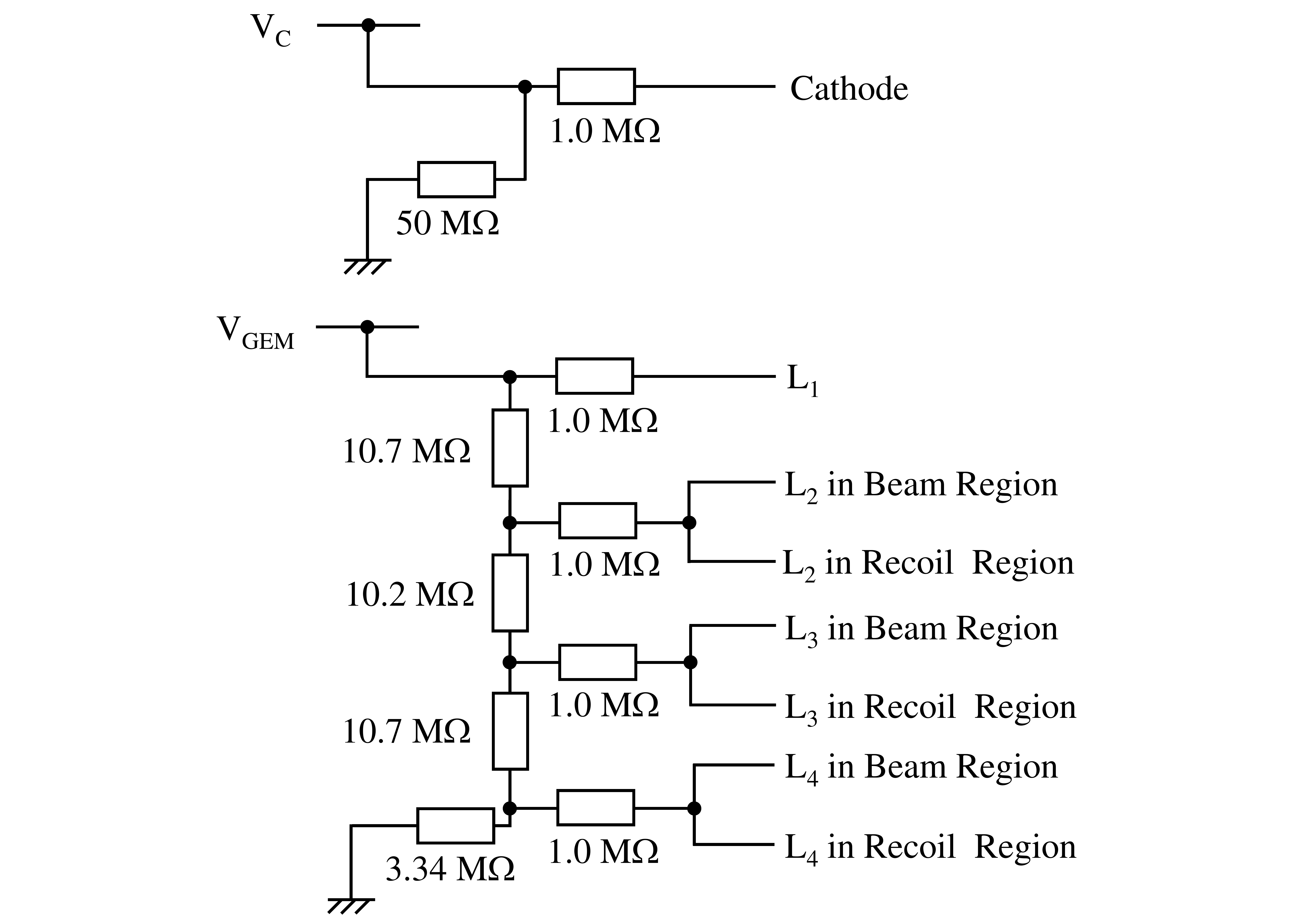}
		\caption{Resistor configuration to supply biases to cathode plate, and electrodes L$_{1}$, L$_{2}$, L$_{3}$ and L$_{4}$.}\label{ResistorSetOffline}
	\end{center}
\end{figure}

Figure \ref{ResistorSetOffline} shows resistor chain to supply biases for the cathode plate and the electrodes of the prototype DG-M-THGEM. 
The bias V$_{\mathrm{C}}$ is for the cathode.  
The bias to each electrode of the prototype DG-M-THGEM was supplied by V$_{\mathrm{GEM}}$ through a resistor divider.
It should be noticed that the beam and recoil regions had the same bias in the present test.
The field strength in the drift region, which was determined by the biases of the cathode plate and L$_{1}$, 
was controled by  combination of V$_{\mathrm{C}}$ and V$_{\mathrm{GEM}}$ in order to  keep to be 1~kV/cm/atm, 
i.e. V$_{\mathrm{GEM}}$ and V$_{\mathrm{C}}$ were simultaneously changed from -1725~V to -1975~V and -2525~V to -2775~V in 25~V steps, respectively.  
The drift velocity of electrons is estimated to be 1~$\mu$s/cm at the present field strength by a simulation program, Garfield++~\cite{Garfield}.
As shown in Fig.~\ref{DGMTHGEM}, the distances between L$_{1}$ - L$_{2}$ and L$_{3}$ - L$_{4}$ are same, 
but one of L$_{2}$ - L$_{3}$ is different from them; 
therefore, the resistor chain was adjusted so that electric fields of L$_{1}$ - L$_{2}$, L$_{2}$ - L$_{3}$ and L$_{3}$ - L$_{4}$ have same strength.

The effective gas gain $G_{\mathrm{eff}}$ is defined as a ratio of $Q_{\mathrm{meas}}$ to $Q_{\mathrm{in}}$, 
\begin{equation}\label{GeffEq}
G_{\mathrm{eff}} = \frac{Q_{\mathrm{meas}}}{Q_{\mathrm{in}}} , 
\end{equation}
where $Q_{\mathrm{meas}}$ is the measured charge from the readout pad and 
$Q_{\mathrm{in}}$ is the initial charge. 
$Q_{\mathrm{in}}$ is calculated by the elementary charge, $e$, stopping power of $\alpha$ particle though the gas, $dE/dx$, mean energy for ion-electron pair creation, $W$, and path length of $\alpha$ particle, $dX$. 
$Q_{\mathrm{in}}$ is described as following,
\begin{equation}\label{Qin_off}
Q_{\mathrm{in}} = \frac{e}{W} \cdot \frac{dE}{dx} \cdot dX , 
\end{equation}
where we used W = 36.5 eV~\cite{Wvalue} for hydrogen gas, and $dX = 15$~mm of the length of the readout pad.
Ambiguity of $dX$ is estimated to be within 0.01~mm, because the $\alpha$ particles were collimated, 
and the angular dispersion is up to  2.3~degree.
Therefore, the $dX$ is assumed to be same as the length of the readout pad.

\begin{figure}[h]
	\begin{center}
		\includegraphics[width=0.8\linewidth, bb=0 0 567 408]{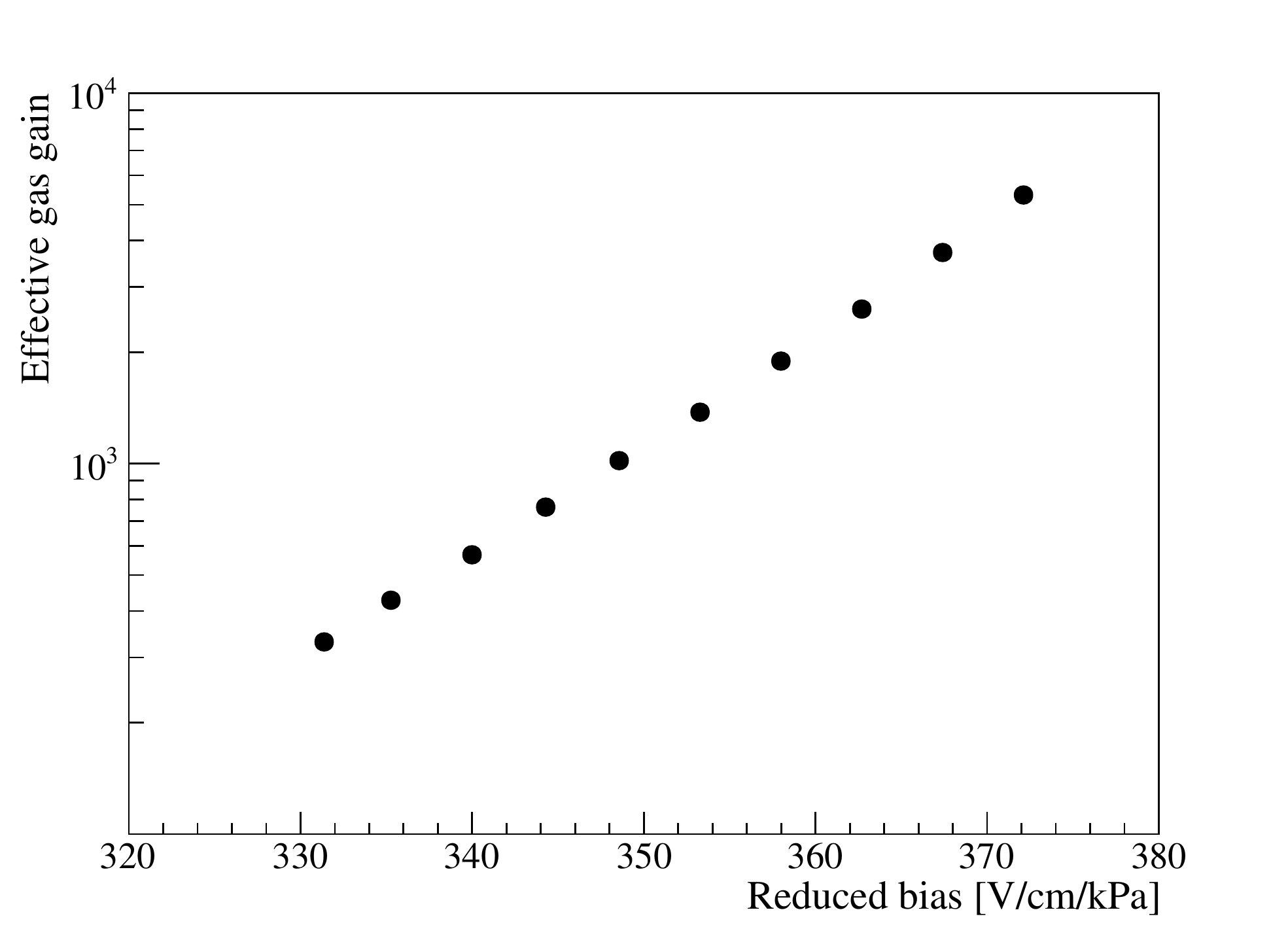}
		\caption{Effective gas gain of prototype DG-M-THGEM as a function of reduced bias between L$_{1}$ and L$_{4}$. 
		The reduced bias is calculated using the gas pressure monitored in each measurement.} \label{GainCurve}
	\end{center}
\end{figure}

Figure~\ref{GainCurve} shows the effective gas gain as a function of reduced bias. 
The reduced bias is derived by dividing the electric field strength between L$_{1}$ and L$_{4}$ with the distance between these electrodes of 0.12~cm and the gas pressure.
The gas pressure was monitored in each measurement and it varied between 39.80~kPa and 40.19~kPa.
Our required effective gas gain for the recoil region is more than $2 \times 10^{3}$, which was achieved with the reduced bias above 360~kV/cm/kPa.
In the present condition, we could not obtain the effective gas gain over $5.31 \times 10^{3}$ due to discharges.
\subsection{Effective gas gain and charge resolution using heavy ion beam}\label{sec3.2}
Measurement of effective gas gain and charge resolution using heavy-ion beam (the program number 15H307) 
was performed at a synchrotron accelerator facility, Heavy Ion Medical Accelerator in Chiba (HIMAC), National Institutes for Quantum Science and Technology (QST).
$^{132}$Xe beam with the energy of 185~MeV/nucleon from the synchrotron accelerator 
was introduced from the end of the beam transport line to experimental setup and injected into the CAT-S.
The repetition time of the synchrotron was 3.3~second. 
So-called slow-extraction mode was applied to have the moderate and uniform intensity for a certain duration. 
The typical extraction duration was 1.5 second.
The number of the $^{132}$Xe particles in each beam pulse was monitored with a diamond detector~\cite{CVDDiaDet} located 1077-mm upstream of the CAT-S. 
The position distribution of the beam was measured with two  low-pressure multi-wire drift chambers (MWDCs)~\cite{LPMWDC} which were installed Z=672~mm and Z=1034~mm downstream of the CAT-S. 
Here we define the beam axis as Z.
the horizontal axis as X, and the vertical axis as Y.
Their origins are set to the center of the CAT-S active area. 

\begin{figure}[h]
	\begin{center}
		\includegraphics[width=1.0\linewidth,bb=0 0 2384 1134]{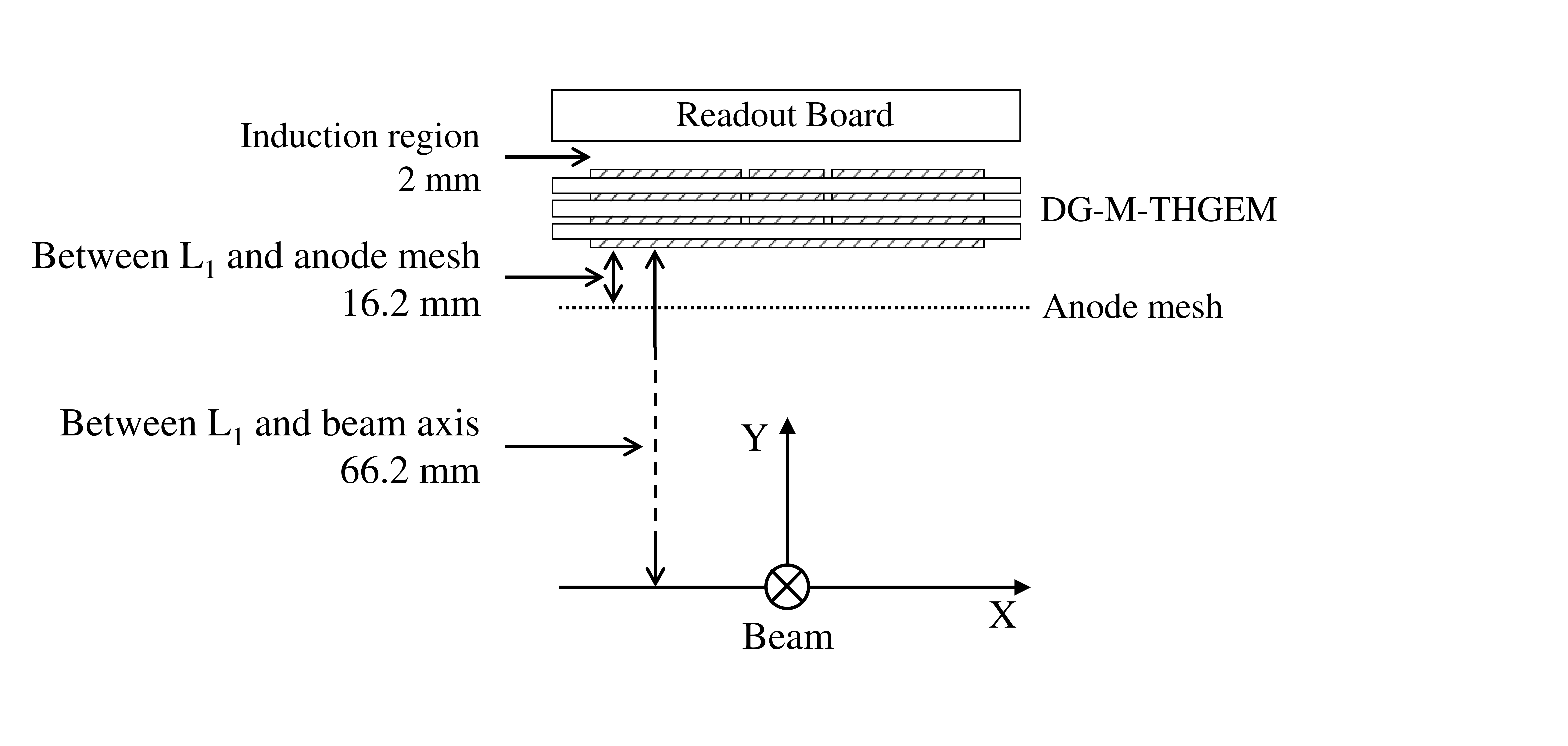}
		\caption{(color online) 
		Schematic view of configuration in the CAT-S chamber for an experiment using $^{132}$Xe beam from upstream. 
		It show the positions of DG-M-THGEM, anode mesh, and readout board relative to beam axis. 
		The origin is the center of the active area of the CAT-S on the beam trajectory.}\label{ExpSetHIMAC}
	\end{center}
\end{figure}

Conﬁguration of the CAT-S with the prototype DG-M-THGEM is shown in Fig.~\ref{ExpSetHIMAC}.
It should be noticed that the configuration is upside down from Fig~\ref{ExpSetOffline}. 
Cathode plate of drift field is not drawn in Fig.~\ref{ExpSetHIMAC}.
The gas pressure and flow rate were 40~kPa and 100~cm$^{3}$/min, respectively.
The oxygen concentration was kept at 0.01\% lower limit of the monitor or less.
The drift field formed by the CAT-S field cage was set to be 1~kV/cm/atm.
Thus, the estimated drift velocity of electron was 1~$\mu$s/cm, as was the case with the test bench experiment.
The electrons drift toward +Y direction.
An anode mesh was installed at 16.2~mm below L$_{1}$. 
The anode mesh is made of SUS304. The diameter and the pitch of the wires of the anode mesh are 30~$\mu$m and 254~$\mu$m, respectively.
The electric field strength between the anode mesh and L$_{1}$ was 2.39~kV/cm/atm.
Readout electrode was mounted at 2~mm above L$_{4}$. 
There are 416 readout pads, which has an equilateral triangle shape of 7-mm side formed on PCB. 
The beam was injected at  66.2~mm below L$_{1}$ in the beam region. 
Each readout pad was connected to the preamplifier RPA-211.
The output signals of the preamplifier were digitized by V1740 flash Analoge-to-Degital convertor (FADC), CAEN Co. Ltd. 
The sampling rate was  50~MHz.
The total charge read by the pad was obtained by summing up the samples.
The obtained total charge was calibrated using the pulse pulser module and the charge injector.

\begin{figure}[h]
	\begin{center}
		\includegraphics[width=1.0\linewidth,bb=0 0 1984 1276]{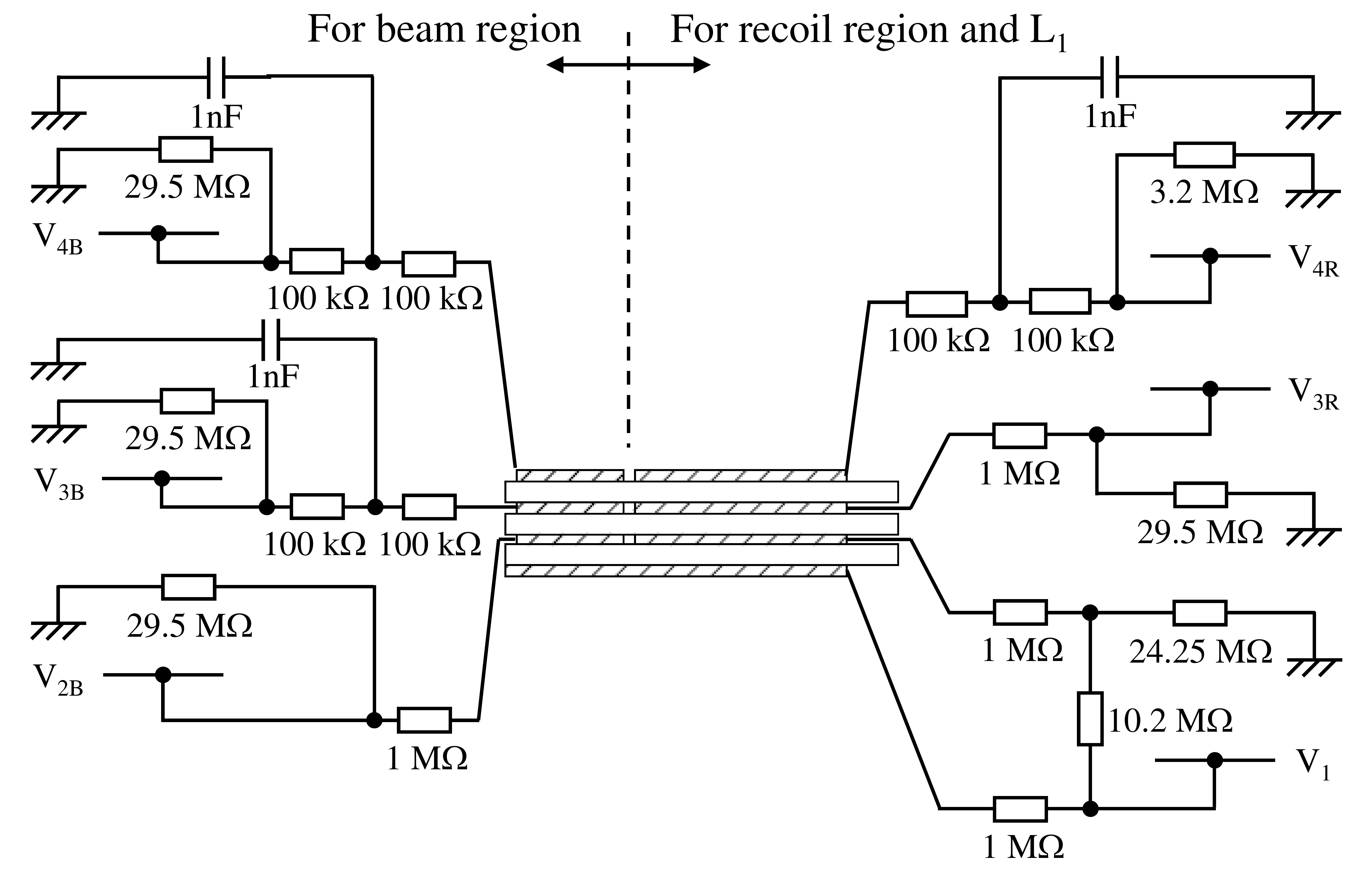}
		\caption{Resistor conﬁguration to supply voltage to electrodes L$_{1}$, L$_{2}$ , L$_{3}$ and L$_{4}$ of DG-M-THGEM shown in Fig~\ref{DGMTHGEM}.
		The electrodes are L$_{1}$, L$_{2}$, L$_{3}$, and L$_{4}$ from the bottom of this figure. The right and left sides are for recoil and beam regions, respectively. 
		V$_{1}$ supplyed bias to L$_{1}$ and L$_{2}$ of the recoil region with resistor chain. 
		To the other electrodes, V$_{\mathrm{3R}}$, V$_{\mathrm{4R}}$, V$_{\mathrm{2B}}$, V$_{\mathrm{3B}}$, and V$_{\mathrm{4B}}$ supplyed biases 
		to L$_{3}$ and L$_{4}$ of the recoil region, and L$_{2}$, L$_{3}$, L$_{4}$ of the beam region, respectively.} \label{ResistorSetHIMAC}
	\end{center}
\end{figure}

Figure~\ref{ResistorSetHIMAC} shows resistor conﬁguration to supply biases to the electrodes of the prototype DG-M-THGEM.  
The biases to L$_{1}$ and L$_{2}$ of the recoil region were supplied by V$_{1}$ through the resistor chains. 
The biases to other electrodes, L$_{3}$ and L$_{4}$ of the recoil region, and L$_{2}$, L$_{3}$, L$_{4}$ of the beam region, 
were supplied by V$_{\mathrm{3R}}$, V$_{\mathrm{4R}}$, V$_{\mathrm{2B}}$, V$_{\mathrm{3B}}$, and V$_{\mathrm{4B}}$, respectively.
Consequently, the gas gain of the beam region and the recoil region can be controlled independently.


The definition of the effective gas gain is the same as Eq. (\ref{GeffEq}). 
The $Q_{\mathrm{in}}$ induced by the $^{132}$Xe beam was estimated similarily from the stopping power $dE/dx$ which was calculated with LISE++ but with considering escape energy carried by delta rays. 
If the delta rays escape from a region of interest, the measured energy deposit become lower than energy loss. 
The energy carried by the delta-rays was estimated using GEANT4~\cite{Geant4}.
Considering both results of simulation and measurement, it is turned out that the ratio of the escaped energy relevant to the energy loss, $\epsilon_{\mathrm{e}}$, is  18.7\% for the $^{132}$Xe beam.
In addition, charges measured by each readout pad are modified from the energy deposit due to diffusion effect in the drift region.
The ratio of the charge measured by each readout pad to the energy deposit, $\epsilon_{\mathrm{d}}$, was estimated by fitting the calculated charges in the readout pads along the trajectory to the measured charges.
The $Q_{\mathrm{in}}$ is modified from Eq.~\ref{Qin_off} as  following, 
\begin{equation}\label{EffEnergyDeposit}
Q_{\mathrm{in}} = \frac{e}{W} \cdot \frac{dE}{dx} \cdot \epsilon_{\mathrm{d}} \cdot (1-\epsilon_{\mathrm{e}}) \cdot dX. 
\end{equation}
Note that the $\epsilon_{e}$ is expected to zero for the low-energy alpha particles. 
The $\epsilon_{d}$ is expected to be one when the readout pad has translational symmetry 
and the change of the stopping power is small enough comparing to the energy deposit.

\begin{figure}[h]
	\begin{center}
		\includegraphics[width=0.8\linewidth,bb=0 0 567 408]{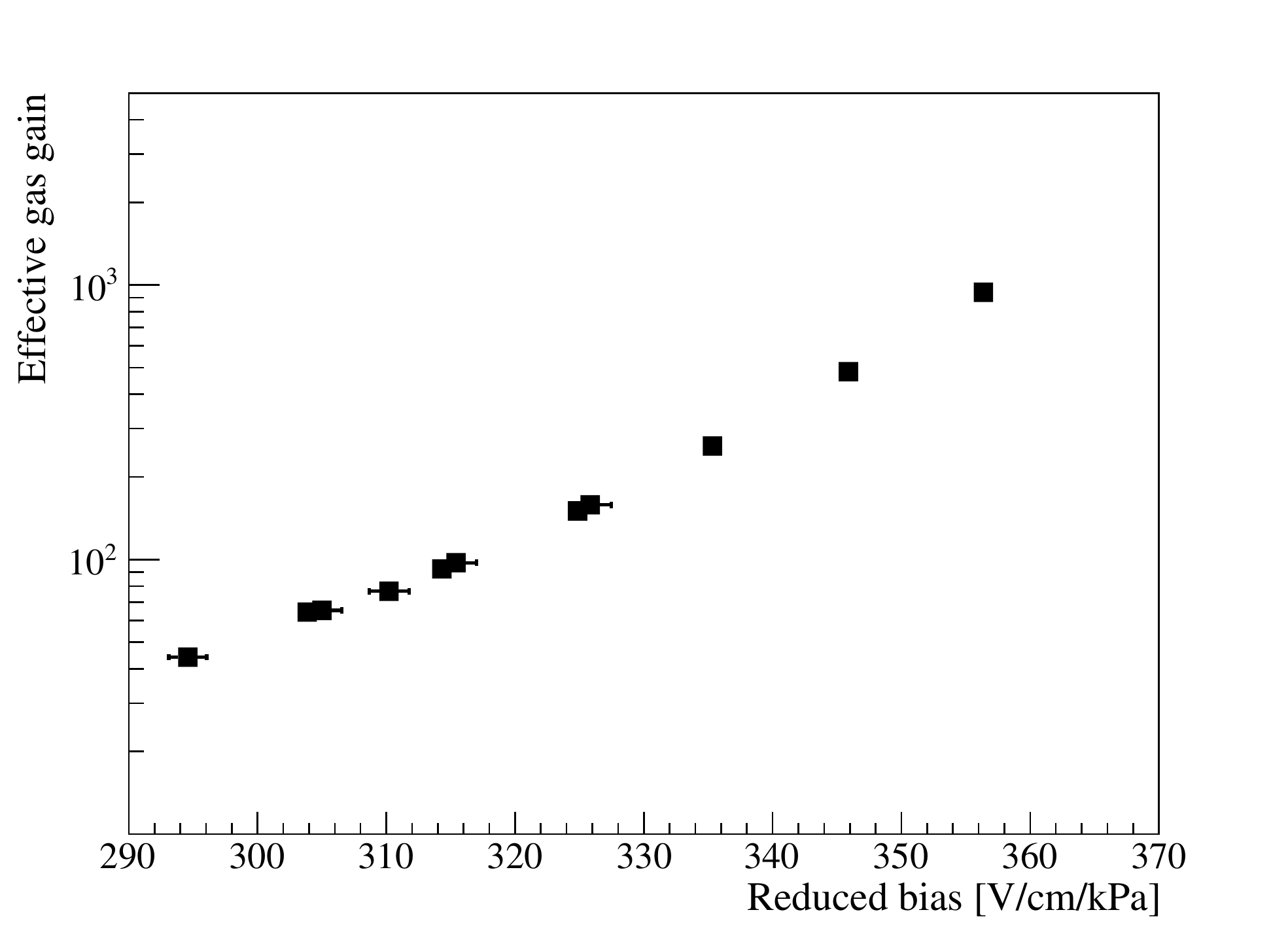}
		\caption{Effective gas gain as a function of the reduced bias with low-intensity $^{132}$Xe beam. 
		The reduced bias is calculated using the gas pressure monitored in each measurement.} 
		\label{GainBeamRecoil}
	\end{center}
\end{figure}

The effective gas gain in the beam region was measured by varying biases. 
Figure~\ref{GainBeamRecoil} shows the measured effective gas gain as a function of the reduced biases with the low-intensity beam of $5 \times 10^{3}$~particles per pulse. 
The effective gas gains were derived by averaging over all of the 52 readout pads in the beam region.
The bias setting is summarized in Table~\ref{BeamVoltageTable}. 
It was reported that lower bias applying to first multiplier stage, which corresponds between L$_{1}$ and L$_{2}$, can control ion backflow~\cite{NSCLMTHGEM}.
We tried to apply slightly lower bias to final multiplier stage of L$_{3}$-L$_{4}$ in the beam region by varying only V$_{\mathrm{4B}}$ while fixing V$_{1}$, V$_{\mathrm{2B}}$ and V$_{\mathrm{3B}}$, aiming to supress the ion back-ﬂow.
The biases of the recoil region, V$_{\mathrm{3R}}$ and V$_{\mathrm{4R}}$ were fixed to make electric fields between each electrodes have same strength.
The effective gas gain lower than $1 \times 10^{2}$ can be achieved in the beam region while the effective gas gain of $2 \times 10^{3}$ at the recoil region was hold. 
The effect of the high-intensity beam injection will be discussed in the next subsection.

\begin{table}[!h]
\caption{Bias setting between each electrodes in beam and recoil region for experiment with $^{132}$Xe beam.}
\label{BeamVoltageTable}
\centering
\begin{tabular}{l c c}
\hline
 & Beam region & Recoil region \\ 
\hline
L$_{1}$ - L$_{2}$ & 596 V & 596 V \\
L$_{2}$ - L$_{3}$ & 570 V & 570 V \\
L$_{3}$ - L$_{4}$ & 548 - 248 V & 598 V \\
\hline
\end{tabular}
\end{table}

\begin{figure}[h]
	\begin{center}
		\includegraphics[width=1.0\linewidth,bb=0 0 2551 1417]{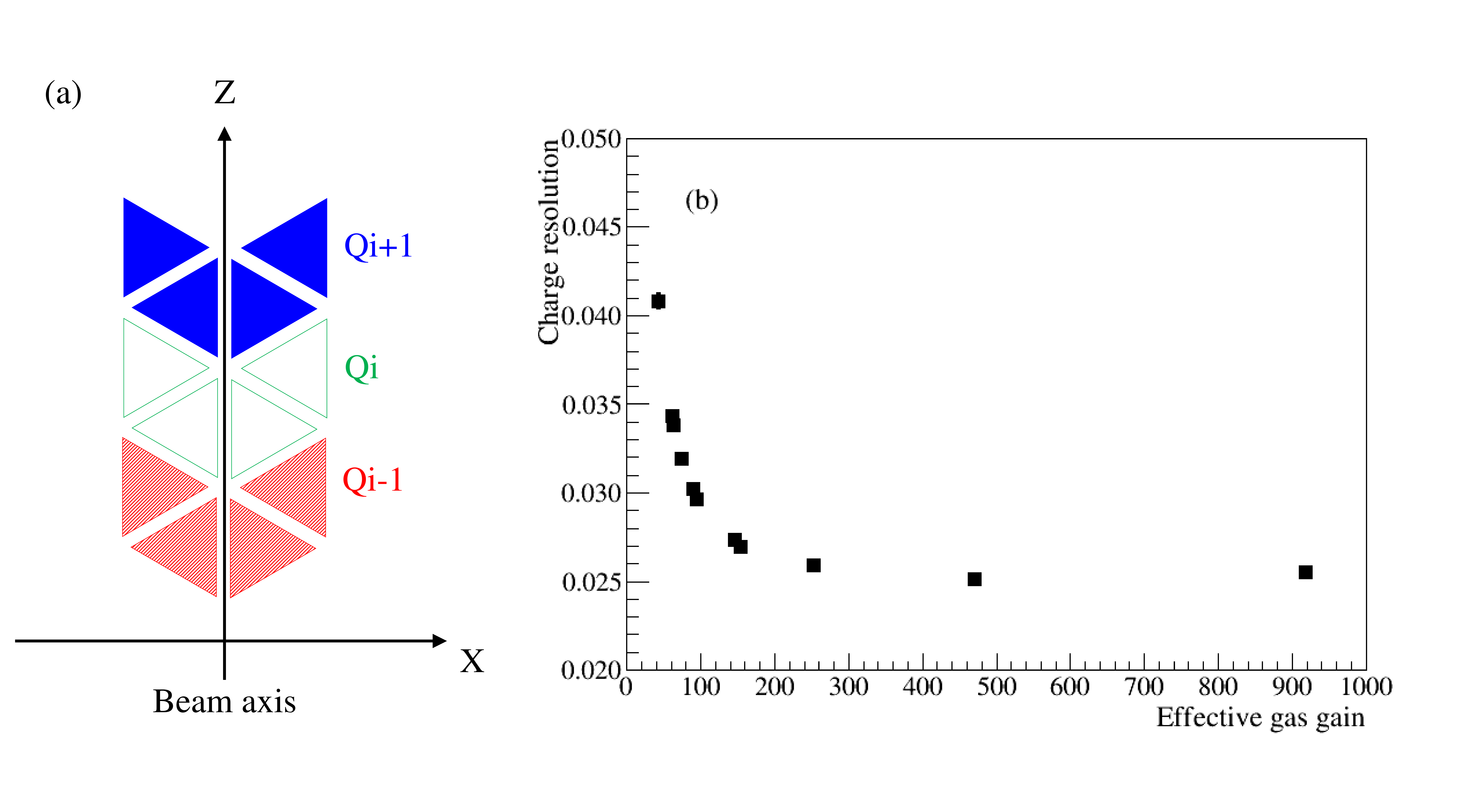}
		\caption{(color online) (a) Definition of pad groups to evaluate charge resolutions.  
		Four triangles with same color are in one group.
		(b) Charge resolution as a function of effective gas gain. 
		}\label{ResVSGain}
	\end{center}
\end{figure}

Charge resolution was also derived with the same bias setting of Table~\ref{BeamVoltageTable}. 
In the present analysis, the charge resolution defined as following; pads were formed in a group defined as shown in Fig.~\ref{ResVSGain}~(a).
Neighboring four triangles are in one group. 
The collected charge by $i$-th group is denoted by $Q_{i}$. 
Index, $i$, is ordered along Z axis, from 0 to 12.
All the $Q_{i}$ is expected to be the same for all $i$-th groups, because the beam energy was sufficiently high and its stopping power was constant at 94.3~keV/mm within 0.2\% overall the active area. 
Thus, we defined a residual of $Q_{i}$ as following, 
%
\begin{equation}\label{EqChargeRes}
\Delta Q = Q_{i} - \frac{(Q_{i-1}+Q_{i+1})}{2}.
\end{equation}
%

The charge resolutions as a function of the effective gas gain are shown in Fig.~\ref{ResVSGain}~(b). 
The charge resolutions were derived averaging of the 11 groups along Z axis.
The charge resolutions significantly depend on the effective gas gain.  
This indicates that the charge resolution is mainly determined by the statistics of the number of amplified electrons.
The charge resolution much smaller than 10\% is achieved over all effective-gas-gain region.

\subsection{Test with high-intensity beam}\label{sec3.3}
In this subsection, sensitivity of the effective gas gain for the beam intensity was discussed 
assuming the bean intensity dependence of the effective gas gain and the initial charge, 
as described in the following equation which modified from the Eq.~(\ref{GeffEq}), 
\begin{equation}\label{QmeasBIDepEq}
Q_{\mathrm{meas}} = G_{\mathrm{eff}}(I_{\mathrm{beam}}) \times Q_{\mathrm{in}}(I_{\mathrm{beam}}).
\end{equation}
Therefore we consider that the $Q_{\mathrm{in}}$ and $G_{\mathrm{eff}}$ are treated as a function of the beam intensity, $I_{\mathrm{beam}}$. 
Ideally the $Q_{\mathrm{meas}}$ should not be changed by $I_{\mathrm{beam}}$, but it has sensitivities for $I_{\mathrm{beam}}$ through this equation. 

\begin{figure}[h]
	\begin{center}
		\includegraphics[width=0.8\linewidth,bb=0 0 567 408]{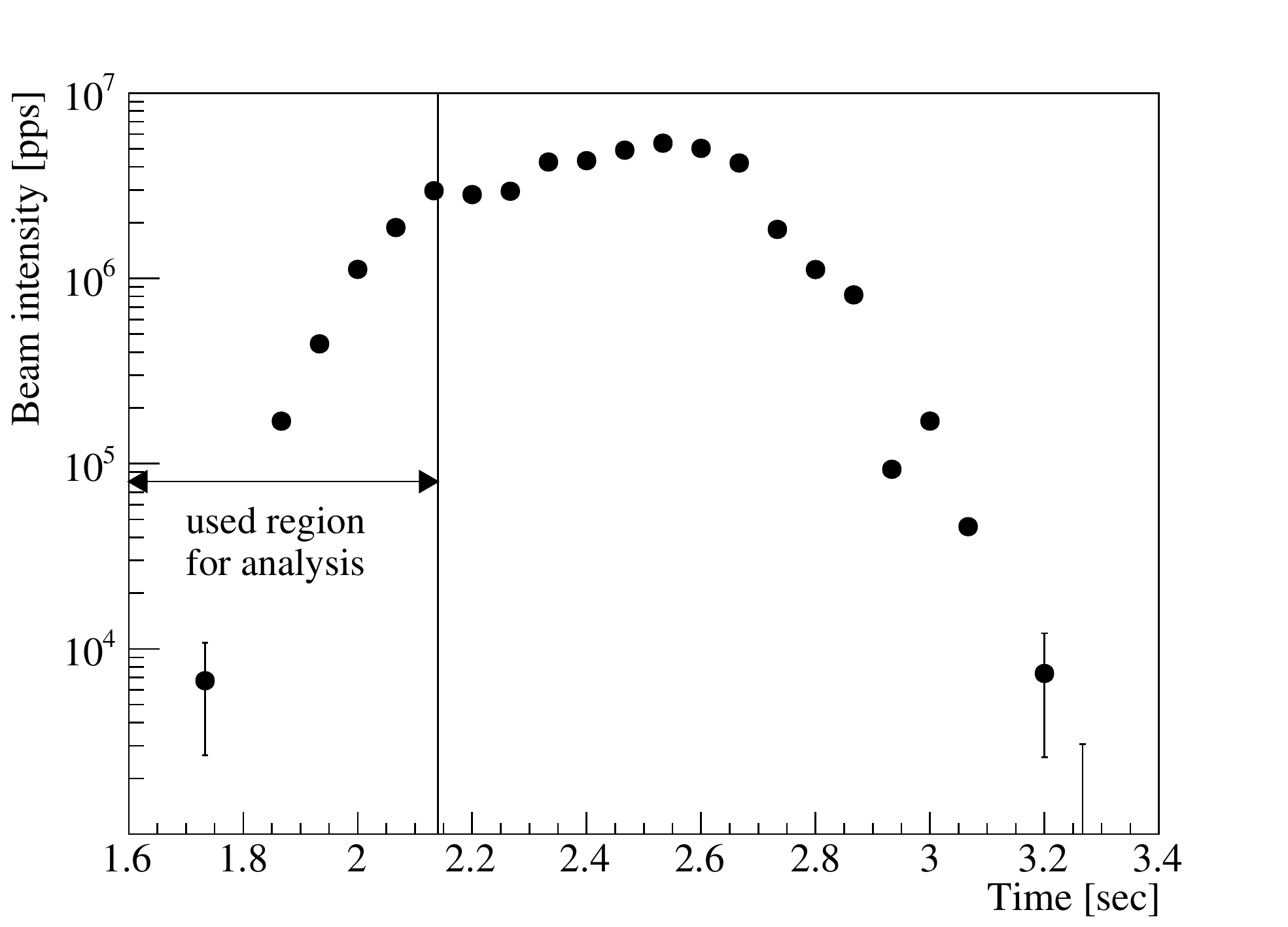}
		\caption{Beam intensity distribution as a function of time in one beam duration when integrated intensity was $1 \times 10^{6}$~particles per pulse.
		The origin of time is arbitrary. 
		\label{BeamStruct}}
      \end{center}
\end{figure}

The beam intensity has a time structure within the extraction duration. 
Figure~\ref{BeamStruct} shows the beam intensity distributions within the extraction duration when the integrated intensity was $1 \times 10^{6}$~particles per pulse.
Horizontal axis shows the time in one beam duration and the origin of the time is arbitrary. 
Vertical axis is averaged beam intensity in each time bin.
The maximum beam intensity is larger than the average of the total number of the beam in one duration.
Considering the beam intensity derived in each time bin, we evaluated the beam intensity dependence of the effective gas gain. 
Table~\ref{BISpillTable} shows the average beam intensity for each time bin considered in the evaluation.

\begin{table}[!h]
\caption{Average beam intensity for each time bin used in Fig.~\ref{BeamStruct}.}
\label{BISpillTable}
\centering
\begin{tabular}{c c}
\hline
time range [s] & average beam intensity [$10^{3}$~pps] \\ 
\hline
1.60 - 1.78 & 1.0 \\ 
1.84 - 1.86 & 120 \\
1.86 - 1.88 & 220 \\
1.88 - 1.91 & 440 \\
1.91 - 1.95 & 660 \\
1.95 - 2.14 & 2500 \\
\hline
\end{tabular}
\end{table}

\begin{figure}[h]
	\begin{center}
		\includegraphics[width=0.6\linewidth,bb=0 0 567 550]{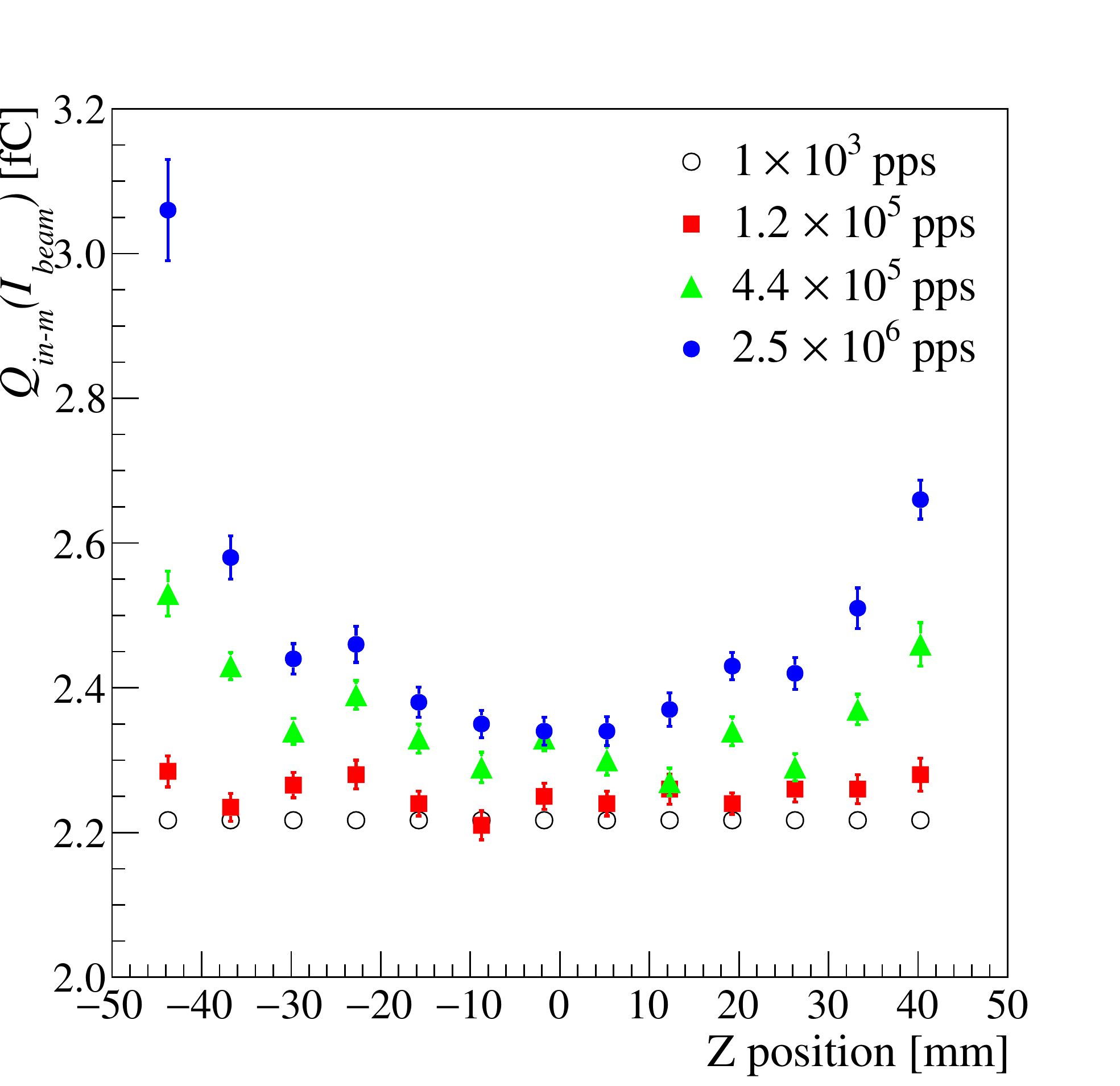}
		\caption{(color online)  $Q_{\mathrm{in-m}}(I_{beam})$ distribution as a function of Z, where $Q_{\mathrm{in-m}}(I_{beam})$ depends on beam intensity also. 
		} \label{QZDist}
      \end{center}
\end{figure}

Assuming that $G_{\mathrm{eff}}(I_{\mathrm{beam}})$ was not so sensitive for the beam intensity, i.e. $G_{\mathrm{eff}}(I_{\mathrm{beam}}) = G_{\mathrm{nom}}$,  
we derive $Q_{\mathrm{in-m}}(I_{\mathrm{beam}}) = Q_{\mathrm{meas}}/G_{\mathrm{nom}}$ as shown in Fig.~\ref{QZDist}.
Here we presumed the nominal effective gas gain, $G_{\mathrm{nom}} = 76.7$, was the effective gas gain obtained in low-intensity $^{132}$Xe beam experiment 
at the supplied bias of 323~V between L$_{3}$ and L$_{4}$ in the beam region. 
Figure~\ref{QZDist} shows $Q_{\mathrm{in-m}}(I_{\mathrm{beam}})$ depends not only on the beam intensity but also the position along the beam axis (Z).
As the beam intensity increases, $Q_{\mathrm{in-m}}(I_{\mathrm{beam}})$ increases overall Z. 
The distribution has a parabolic shape with a symmetry at Z = 0.

\begin{figure}[h]
	\begin{center}
		\includegraphics[width=0.6\linewidth,bb=0 0 567 550]{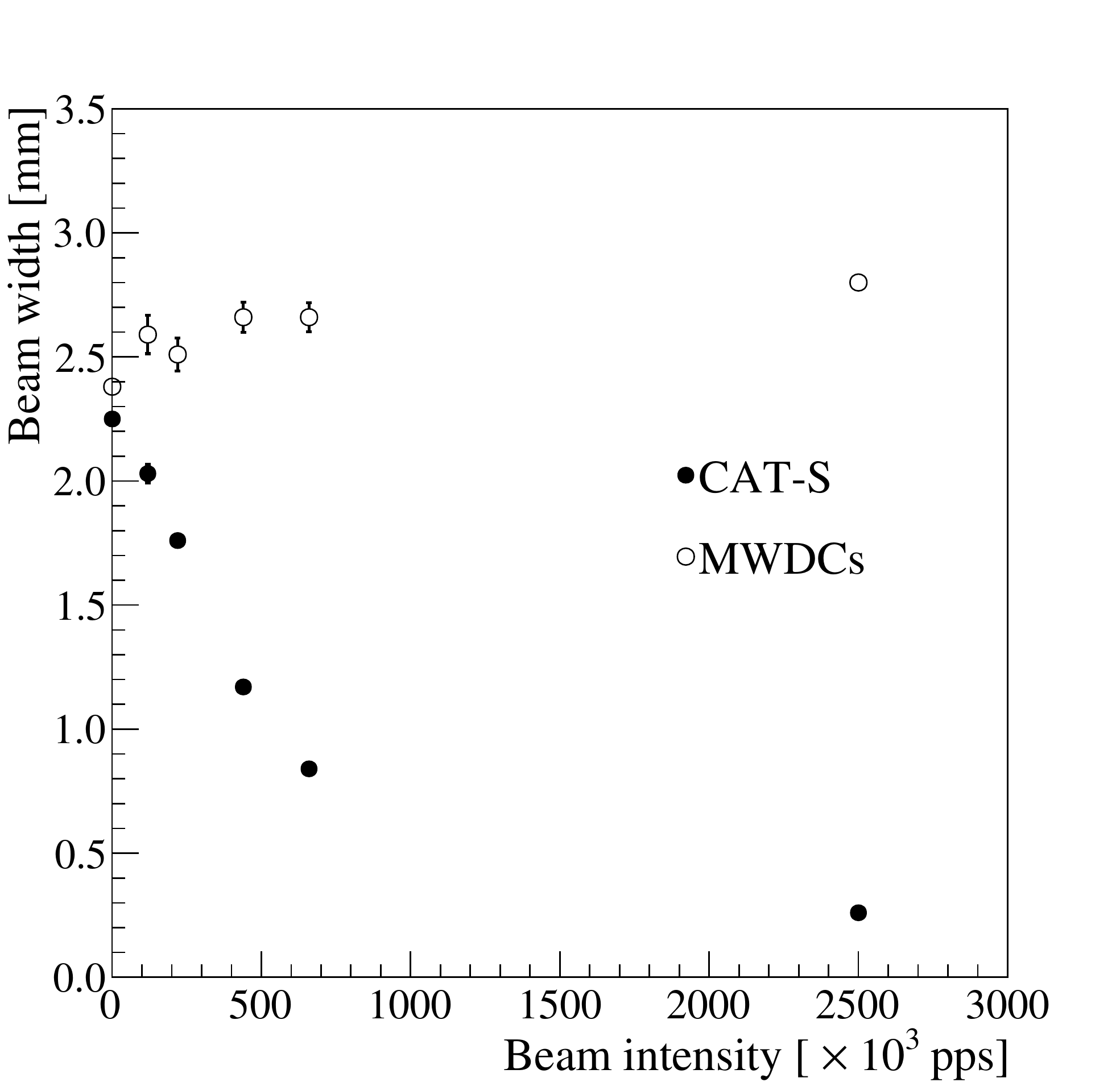}	
		\caption{Beam profile on X axis derived by CAT-S and MWDCs. 
		The beam width measured by the CAT-S was shrinked due to effect of ion pillar with beam intensity increment, while no effect is shown by the MWDCs.}
\label{BeamWidth}
      \end{center}
\end{figure}

To explain this parabolic shape, we employed ``ion pillar model''~\cite{IonPillar}.
The ion pillar model has been proposed 
to evaluate electric field distortion in the drift region caused by the ion-backflow from GEMs.
The backflow ions move from the GEMs to the cathode, 
then a pillar of positive ions along the beam trajectory is formed. 
In the previous studies, electric field distortion in the drift direction (Y direction) has been discussed, 
but similar electric field distortion also occurs along the beam axis (Z direction) and its vertical direction (X direction).
The ions attract drifting electrons from outside the beam region, consequently $Q_{\mathrm{in-m}}(I_{\mathrm{beam}})$ increases in the beam region.
The number of the ions increases with increment of the beam intensity, the behavior of $Q_{\mathrm{in-m}}(I_{\mathrm{beam}})$ in Fig.~\ref{QZDist} can be explained.
Figure~\ref{BeamWidth} shows the comparison of beam profile on the X axis which is measured by the CAT-S and the MWDCs.
The beam profile measured by the MWDCs shows the similar size independent of the beam intensity. 
However, ones measured by the CAT-S was observed to become narrow according the beam intensity, 
because the attracted electrons converge on the beam axis.

The effect of ions in the pillar on electrons which reach to L$_{1}$ was investigated by electron-drift  simulation.
The ion density in the pillar was assumed to be distributed homogeneously in the Z and Y direction and Gaussian in the X direction.
The beam was injected 5~cm below L$_{1}$. 
The distance from L$_{1}$ to the cathode plane of the field cage was approximately 250~mm.
The drift velocities of a hydrogen ion and an electron were 0.01~cm/$\mu$s and 1.2~cm/$\mu$s, respectively.
Thus the time for the back-flow ions to reach the cathode was approximately 3~ms. 
As shown in Table~\ref{BISpillTable}, the time bin for each beam intensity was from about 20~ms to 200~ms.
Therefore, the number of the ions in the pillar was considered to be immediately saturated depending on the beam intensity within the time bin.
The number of the electrons reaching to L$_{1}$ along the beam trajectory, $Q_{\mathrm{in-s}}(I_{\mathrm{beam}})$, was derived by the electron-drift simulation with Garfield++~\cite{Garfield} while varying the amount of ions in the pillar.

\begin{figure}[h]
	\begin{center}
		\includegraphics[width=1.0\linewidth, bb=0 0 567 553]{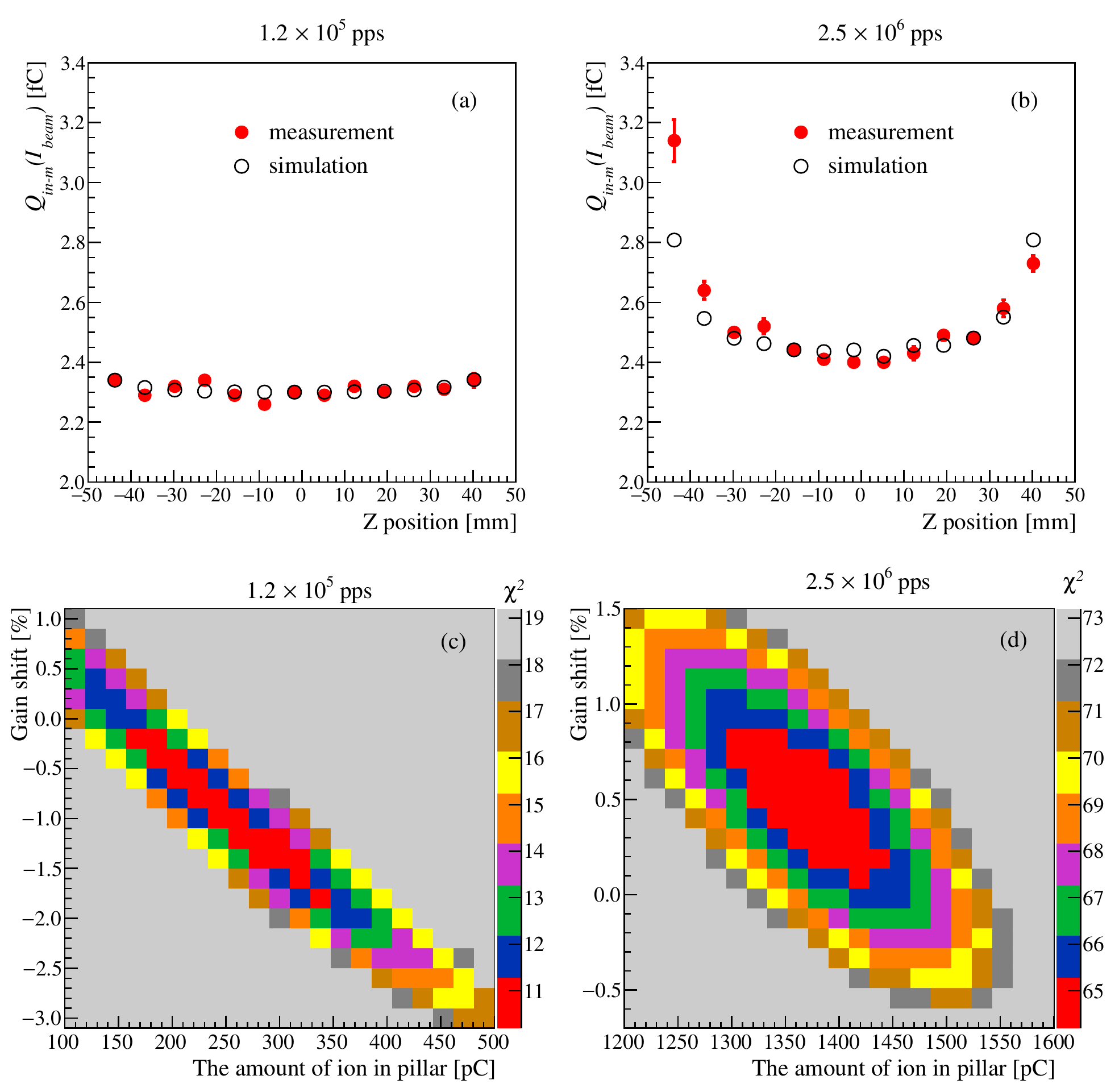}
		\caption{(color online) (a) and (b): Position distributions of $Q_{\mathrm{in-m}}(I_{beam})$, 
				which is the number of electrons reaching electrode L$_{1}$ of DG-M-THGEM, along beam trajectory Z 
				at beam intensity of $1.2 \times 10^{5}$ and $2.5 \times 10^{6}$~pps, respectively. 
				Solid circles and open circles are measured value and simulated results at $\chi^{2}$ minimum by the least square ﬁtting, respectively.
				(c) and (d): $\chi^{2}$ distributions as a function of the amount of ions in pillar and effective-gas-gain shift at the beam intensity of $1.2 \times 10^{5}$ and $2.5 \times 10^{6}$~pps, respectively.}
				 \label{QZChi}
      \end{center}
\end{figure}

To reproduce $Q_{\mathrm{in-m}}(I_{\mathrm{beam}})$ in Fig.\ref{QZDist}, 
we explain $Q_{\mathrm{in-m}}(I_{\mathrm{beam}})$ using $Q_{\mathrm{in-s}}(I_{\mathrm{beam}})$ as following, 
\begin{equation}\label{QmeasFittingEq}
Q_{\mathrm{in-m}}(I_{\mathrm{beam}}) = \frac{Q_{\mathrm{meas}}}{G_{\mathrm{nom}}} = \frac{Q_{\mathrm{in-s}}(I_{\mathrm{beam}}) \times G_{\mathrm{eff}}(I_{\mathrm{beam}})}{G_{\mathrm{nom}}}. 
\end{equation}
We set $Q_{\mathrm{in-s}}(I_{\mathrm{beam}})$ and $G_{\mathrm{eff}}(I_{\mathrm{beam}})$ as the fitting parameters in least-square fitting method to reproduce the $Q_{\mathrm{in-m}}(I_{\mathrm{beam}})$ in Fig.\ref{QZDist}.
Because $Q_{\mathrm{in-s}}(I_{\mathrm{beam}})$ is depending on the amount of the ion in the pillar, 
the actual parameters to be varied are the amount of ions in the pillar and the effective gas gain. 
Figure~\ref{QZChi} (c) and (d) shows the fitting results of the distributions of the reduced $\chi ^{2}$ as a function of the amount of the ion in the pillar and effective-gas-gain shift for the beam intensities of $1.2 \times 10^{5}$ and $2.5 \times 10^{6}$~pps, respectively.
Here, the effective-gas-gain shift is defined as $\{G_{\mathrm{eff}}(I_{\mathrm{beam}})-G_{\mathrm{nom}}\}/G_{\mathrm{nom}}$.
Searching $\chi ^{2}$ minimum point on  Fig.~\ref{QZChi} (c), the number of the ion in the pillar and the effective-gas-gain shift are $2.6 \times 10^{2}$~pC and -1~\%, respectively, 
Using these parameters, we can reproduce  $Q_{\mathrm{in-m}}(I_{\mathrm{beam}})$ spectrum as shown Fig.~\ref{QZChi} (a).
Figure~\ref{QZChi} (b) is also reproduced by $1.38 \times 10^{3}$~pC and and 0.4\%.
This result shows that the ion backflow from the DG-M-THGEM is significantly large to show its effect.
The required effective gas gain in the beam region was achieved with the dual gain operation by the lower bias application between L$_{3}$ and L$_{4}$ in the beam region, 
however, the ion backflow effect was not sufficiently small.

\begin{figure}[h]
	\begin{center}
		\includegraphics[width=1.0\linewidth,bb=0 0 567 269]{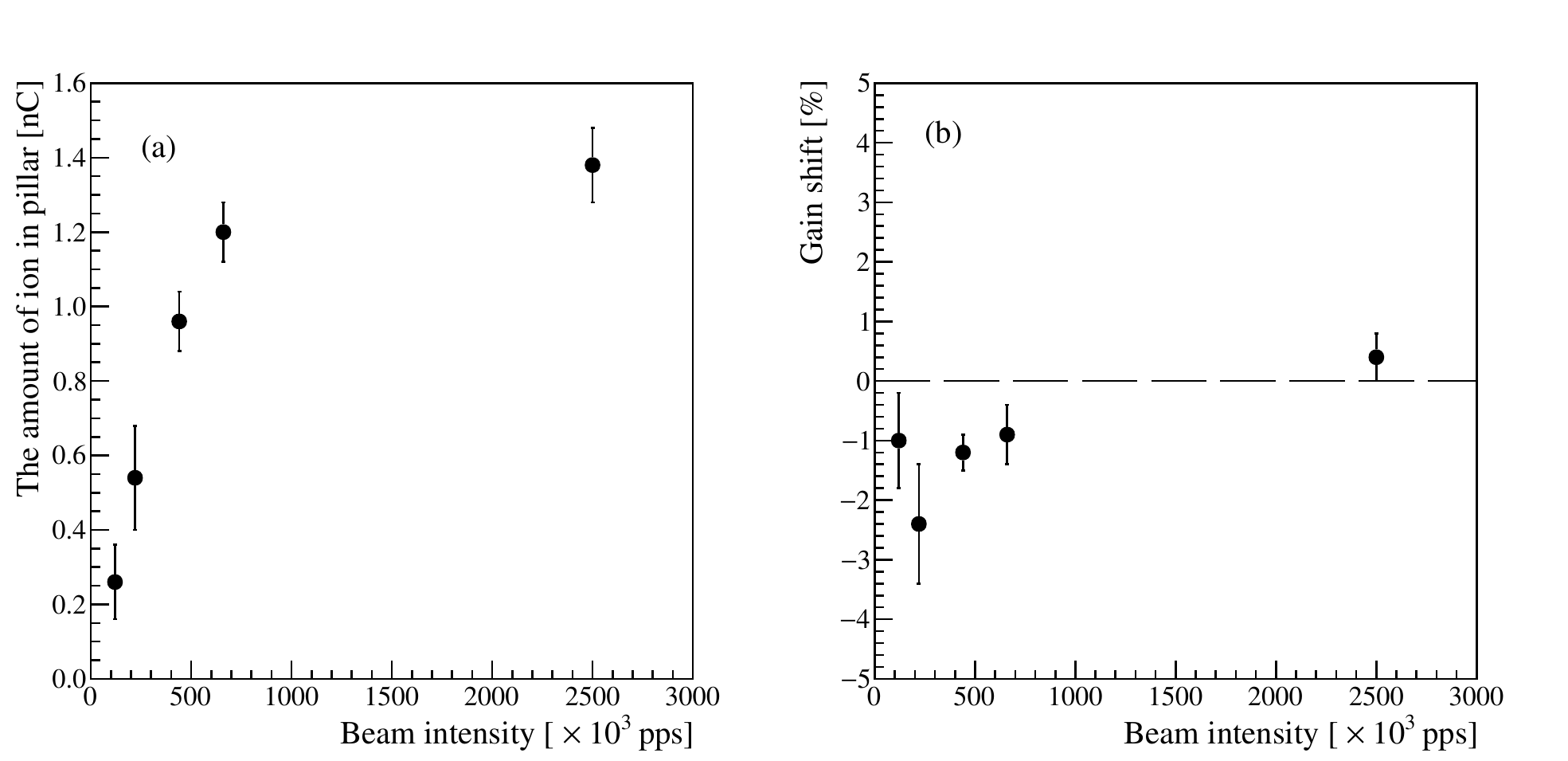}
		\caption{(a) The amount of ions in pillar as a function of beam intensity. (b) Effective-gas-gain shift as a function of beam intensity.} \label{GainShiftAndPillarCharge}
      \end{center}
\end{figure}

Figure~\ref{GainShiftAndPillarCharge} shows the beam intensity dependence of the amount of ion in the pillar and the effective-gas-gain shift. 
Comparing the beam intensity of $2.5 \times 10^{6}$~pps with $1.2 \times 10^{5}$,
its fluctuation is suppressed within 3\% while the amount of ion increases by seven times.
It was found that the amount of ion in the pillar was saturated above the beam intensity of $6 \times 10^{5}$~pps.
It is considered due to decreasing the ion backflow from the DG-M-THGEM as the amount of ions in the pillar increases above a certain level, 
similar to the trend shown in a previous study~\cite{IBFvsSCD}.
The fluctuation of the effective gas gain was sufficiently suppressed to the same level as the charge resolution at the effective gas gain of $1 \times 10^{2}$. 
As shown in Fig.~\ref{BeamWidth}, the electrons attracted by the ion pillar to the beam center; 
therefore electron density increased by a factor of 8.5 at $2.5 \times 10^{6}$~pps comparing to at $1 \times 10^{3}$~pps beam intensity.
It can be explained that the fluctuations were suppressed to 3\% even under conditions where ionized electrons of 6~pc/cm$^{2}$/s flow to the prototype DG-M-THGEM.


\section{Summary}\label{sec5}
The prototype Dual Gain Multilayer Thick GEM (DG-M-THGEM) with the active area of 10~cm $\times$ 10~cm was produced and its performance was evaluated.
The electrodes of multilayer thick GEM, which has an alternating structure of electrodes and insulators, were segmented to three regions. 
The center region and both sides can be applied biases independently to control gas gains individually. 
The performance of the prototype DG-M-THGEM was evaluated in the hydrogen gas at the pressure of 40~kPa. 
The effective gas gains as a function of the reduced bias applied DG-M-THGEM were measured using the $\alpha$ source of $^{241}$Am.
The effective gas gain was achieved up to $5.31 \times 10^{3}$.

The effective gas gain and the charge resolutions in the beam region were evaluated using the heavy-ion $^{132}$Xe beam with the energy of 185~MeV/nucleon and intensity from $5 \times 10^{3}$ to $1 \times 10^{6}$~particles per pulse.
The effective gas gain of lower than $1 \times 10^{2}$ was achieved with the charge resolution of smaller than 3~\% in the beam region 
while maintaining the effective gas gain of $2 \times 10^{3}$ in the recoil region.
The effective-gas-gain stability with increasing the beam intensity was also discussed.
As the beam intensity increases, the initial charges become larger, because the ion pillar attracts electrons.
The effect shrinked the beam width measrured by the CAT-S as the beam intensity increase.
Even if increasing the beam intensity from $1.2 \times 10^{5}$~pps and $2.5 \times 10^{6}$~pps, the effective gas gain fluctuate within only 3\% 
while the amount of ion increased by seven times.
The effective-gas-gain fluctuation was suppressed to the same level as the charge resolution at the effective gas gain of $1 \times 10^{2}$.
The number of electrons around the beam center was about 8.5 times larger at $2.5 \times 10^{6}$~pps than at $1 \times 10^{3}$~pps due to the effect to attract electrons by the ion in the pillar. 
It can be explain that the effective gas gain fluctuate within 3\% even under conditions where ionized electrons of 6~pc/cm$^{2}$/s ﬂow to the prototype DG-M-THGEM.
As a future issue, it is necessary to develop a way to reduce the effect of the ion backflow in order to perform accurate tracking analysis of the trajectory.

%
\section*{Acknowledgment}
This work was performed in part as the Research Project with Heavy Ions at NIRS-HIMAC (the program number 15H307).
The present work was supported by JSPS KAKENHI, Grant Numbers JP23740174, JP15H00834, JP16H06003.


\end{document}